\documentclass[aps,prl,twocolumn,amsmath,amssymb,floatfix]{revtex4}
\usepackage{epsfig,psfrag,subfigure}
\usepackage{graphicx,psfrag,subfigure}
\usepackage{color}
\newcommand\beq{\begin{equation}}
\newcommand\eeq{\end{equation}}
\newcommand\bea{\begin{eqnarray}}
\newcommand\eea{\end{eqnarray}}

\begin{document}

\title{Josephson physics mediated by the Mott insulating phase}
\author{Smitha Vishveshwara$^{1}$ and Courtney Lannert$^{2}$}
\affiliation{$^1$ Department of Physics, University of Illinois at Urbana-Champaign, 1110 W.
Green St, Urbana, IL 61801, USA \\ $^{2}$Wellesley College, Wellesley, MA 02481, USA}

\date{\today}

\begin{abstract}
 We investigate the static and dynamic properties of bosonic lattice systems
 in which condensed and Mott insulating phases co-exist due to the presence of a
spatially-varying potential. We formulate a description of these inhomogeneous
systems and calculate the bulk energy at and near equilibrium. We derive the
explicit form of the Josephson coupling between disjoint superfluid regions
separated by Mott insulating regions. We obtain detailed estimates for the
experimentally-realized case of alternating superfluid and Mott insulating
spherical shells in a radially symmetric parabolically-confined cold atom
system.

\end{abstract}
\maketitle

An important and generic situation presented by many-body quantum phenomena is
that of competing states of matter co-existing in spatially-separated regions
within a given system due to the presence of inhomogeneities.
Transport properties of diverse systems such as the quantum Hall system,
metal-insulator compounds, high $T_c$ superconductors, and more recently, cold
atomic gases, are determined by the tunneling of carriers between conducting
regions that are embedded within insulating regions\cite{Shimshoni98}.
Crucial to understanding thermodynamic and transport features of such systems
is the manner in which conducting regions couple to one another through the
insulating regions. Classic examples of Josephson coupling in superconductors
and cold atoms rely on an externally-imposed potential barrier between
condensed regions \cite{Tinkham_book,Pethick_Smith}. Here, on the other hand,
we explore systems of bosons in which condensed (superfluid) regions exhibit
Josephson physics mediated by Mott-insulating regions of the same bosons. This
model should be germane to a diverse range of systems, in particular, granular
superconductors and high $T_c$ materials where Cooper pairs can be treated as
the bosonic degrees of freedom\cite{SuperconductorBosons}, and trapped cold
atoms in optical lattices where the atoms are bosons\cite{MottSfShells}.
Through an explicit description of these phases in terms of microscopic
parameters, we are able to go beyond phenomenological treatments for obtaining
transport co-efficients in these systems\cite{Shimshoni98}.

\begin{figure}[h]
\includegraphics[width=2.5in]{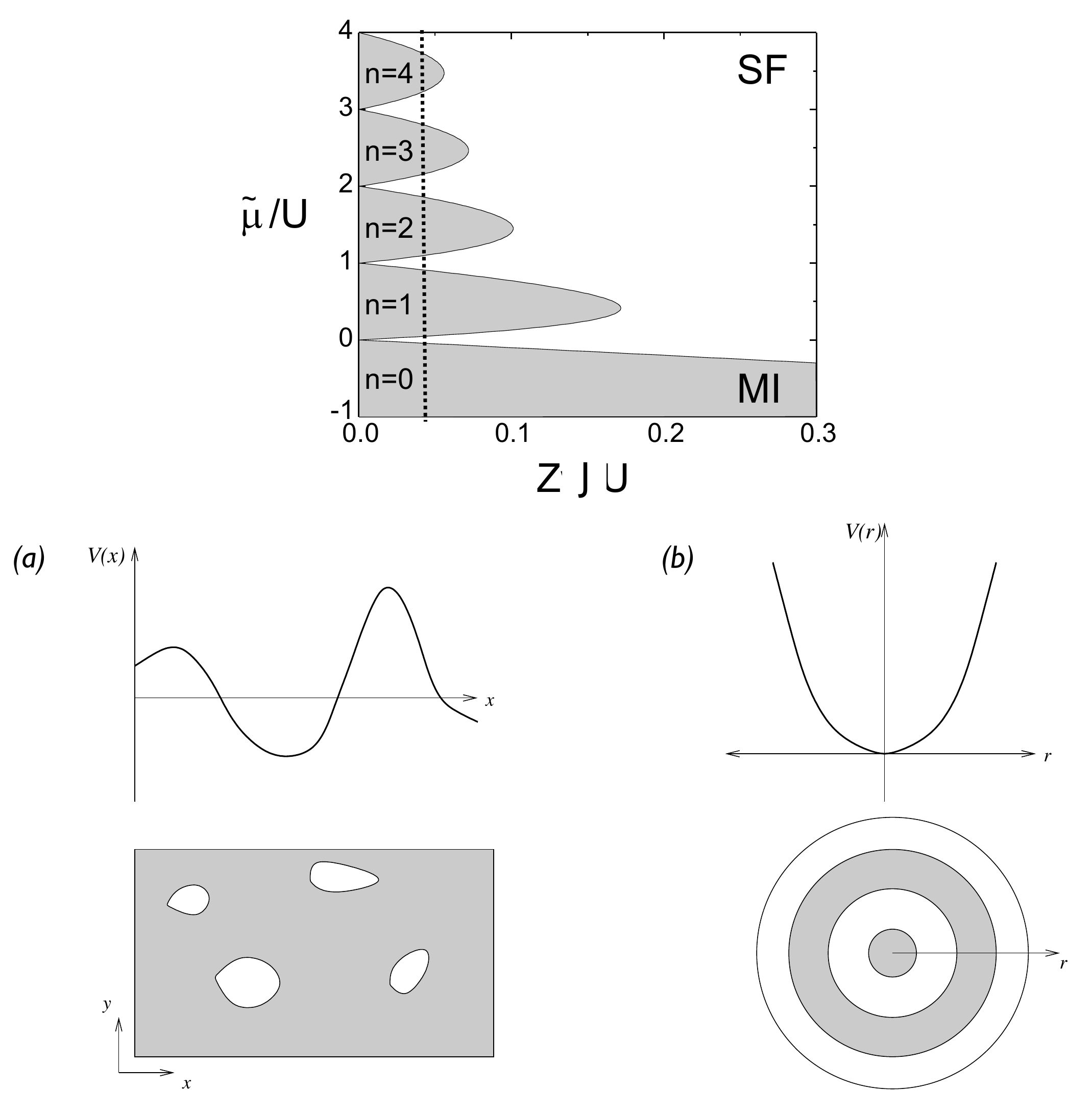}
\caption[]{Top: the zero temperature phases of the Bose-Hubbard model; the dotted line cuts through the phases that could coexist for a fixed small value of $zJ/U$. Below: schematic (a) slowly-varying random potential and (b) harmonic confining potential, and subsequent Mott-superfluid domains.}
\label{fig:domains}
\end{figure}

Towards understanding this physics of co-existent phases, we study a system of
interacting bosons on a lattice in the presence of a smooth potential $V({\bf
r})$ which varies on length scales much larger than the lattice spacing $a$.
Within a local density approximation, the potential is equivalent to a shift in
the local chemical potential $\tilde{\mu}({\bf r})=\mu-V({\bf r})$, where $\mu$
is the global chemical potential determined by the total number of bosons in
the system, $N$. In the situations of interest, shown in Fig.\ref{fig:domains},
the potential $V({\bf r})$ breaks the system into phase-separated domains of
Mott-insulator (wherein interactions pin the number of bosons per site) and of
condensed bosons (which exhibit number fluctuation on each site). In what
follows, we derive the equilibrium properties of the domains, bulk energy
costs for small deviations from equilibrium, dynamics of the condensed regions,
the Josephson coupling between condensed regions mediated by a Mott-insulating
interface, and detailed estimates for the spherically-symmetric situation illustrated in
Fig.\ref{fig:domains}b.

This system can be modeled by the Bose-Hubbard Hamiltonian, describing bosons
whose tunneling between neighboring lattice sites has strength $J$ and whose
on-site repulsive interaction is $U$. For small $J/U$, each
superfluid region is energetically near two Mott insulating phases, say of
occupation $n_0$ and $n_0+1$. To describe this superfluid region, we employ a
pseudo-spin formulation of the Bose-Hubbard
model\cite{Matsubara56,Bruder93,Altman02} that truncates the Hilbert space to
these two occupation numbers on each site. This formulation can be generalized
to include more number states if necessary. But here, for simplicity and as
realized in cold atom systems, we assume that $J/U$ is sufficiently small to
justify the truncation. The two-state Hilbert space maps to a spin-$1/2$ basis
on each site, $i$, with the identifications $|n_0+1\rangle_{i} \leftrightarrow
|\uparrow\rangle_{i}$ and $|n_0\rangle_{i} \leftrightarrow
|\downarrow\rangle_{i}$, where the $|\uparrow/\downarrow\rangle_{i}$ are
eigenstates of the spin operator $s_i^z$ with eigenvalues $\pm 1/2$, and
$b_{i}^{\dagger} = \sqrt{n_0+1}s_{i}^{+}$ ($b_i=\sqrt{n_0+1}s_{i}^{-}$), where
$b_i^{\dagger}$ and $b_i$ denote bosonic creation and annhilation operators,
respectively, on the site $i$. The number operator, $\hat{n}_i=b_i^{\dag}b_i$
is related to the $z$ component of the spin: $\hat{n}_i=n_0+1/2+s_i^z$. With
this mapping, the Hamiltonian takes the form:
\begin{equation}
\label{eq:Ham_spin}
{\cal H}=-J(n_0+1)\sum_{\langle ij\rangle}\left( s^x_i s^x_j+s^y_i
s^y_j\right)+\sum_i (Un_0-\tilde{\mu}_i)s^z_i,
\end{equation}
where $\langle ij\rangle$ denotes a summation over nearest-neighbor sites and
$\tilde{\mu}_i \equiv \tilde{\mu}({\bf r}_i)$. At the mean-field level, to which
we confine ourselves in this Letter, the ground state configuration has the
pseudospins aligned with the local ``magnetic" field, ${\bf B}^0_i= z
J(n_0+1)\left[ 2f^x_i,2f^y_i,\cos\theta_i\right]$, where
$\cos\theta_i=(\tilde{\mu}_i-Un_0)/(zJ(n_0+1))$ with $z$ the coordination
number of the lattice, the fields ${\bf f}_i$ denote expectation values of spin
operators (e.g. $f^z_i=\langle s^z_i \rangle$) and we have assumed ${\bf
f}_i\approx {\bf f}_j$ for nearest-neighbors.
The equilibrium $z$-component
of the pseudospin has the value $f^z_{i0}=(1/2)\cos\theta_i$; the Mott phases
correspond to complete polarization of the pseudospin along the $z$ direction,
i.e. $f^z_{i0}=\pm 1/2$. Within the mean-field approximation,  we can thus
identify the Mott-superfluid boundaries shown in Fig. \ref{fig:domains} as
occurring at the critical values of the external potential $\mu-V({\bf
r}^c_{\pm})=Un_0\pm z J(n_0+1)$, where $\pm$ refers to the boundary at the Mott
$n_0+1$ and $n_0$ phases respectively.

In the condensed phase, a local order parameter can be defined as $\psi =
\langle b^{\dagger}\rangle = \sqrt{n_0+1}f^+$ for $0<f^z\leq 1/2$ and $\psi =
\langle b\rangle = \sqrt{n_0+1}f^-$ for $-1/2<f^z\leq 0$, corresponding to
condensates of particles and holes, respectively. To first order in $J/U$ and
in the continuum limit, the equilibrium order parameter profile (as a function of
$\tilde{\mu} = \mu - V({\bf r})$) follows from the normalization: $f_0^{\pm} = \sqrt{1-{f^z_0}^2}/2$.
Ignoring the energy cost of
variations of $f_0^{\pm}$ from site to site (the Thomas-Fermi approximation),
the order parameter is found to be:
\begin{equation}
\psi({\bf r}) = \sqrt{\frac{z^2J^2(n_0+1)^2-(\tilde{\mu}-n_0U)^2}{4z^2J^2(n_0+1)}}
\label{eq:OP}
\end{equation}
This is of the same form as the Thomas-Fermi order parameter for a traditional
condensate in an external potential $V_{ext}$ and with interaction strength
$g$: $\psi_{TF} = \sqrt{(\mu-V_{ext})/g}$ \cite{Pethick_Smith}. This allows us
to identify the ``effective" confining potential for the superfluid between two
Mott regions in the optical lattice system: $(\mu-V_{ext})_{eff} =
(\tilde{\mu}-n_0U)^2/(zJ[n_0+1])$.
The boson density in the condensed phase is found from
$\langle\hat{n}\rangle = (n_0+1/2)+f^z$ and in equilibrium in the Thomas-Fermi approximation is:
\begin{equation}
\langle \hat{n} \rangle = (n_0+1/2) + \frac{\tilde{\mu}-n_0U}{2zJ(n_0+1)}
\label{eq:density}
\end{equation}
which smoothly interpolates between densities of $n_0+1$ at ${\bf r}_c^+$ and
$n_0$ at ${\bf r}_c^-$.

For mesoscopic superfluid regions, the energy cost for deviations from
equilibrium is non-negligible and is described by the bulk energy $E_B$. Within the Thomas-Fermi approximation, the Hamiltonian, Eq.
(\ref{eq:Ham_spin}), can be expressed in terms of $f^z({\bf r})$:
\begin{equation}
\label{eq:bulk_energy}
E_B(N)=\int\frac{1}{a^3}\left[zJ(n_0+1)({f^z}^2-1)+(n_0U-\tilde{\mu})f^z\right]d{\bf
r},
\end{equation}
where, assuming that variations in the density are over length scales greater
than the lattice spacing, a continuum approximation has been made. In this
approximation, the Mott and superfluid regions are decoupled from one another
and have separate contributions to the bulk energy of the system, $E_B(N)
= E_B^{Mott}(N_M)+E_B^{sf}(N_S)$, where $N_M$ and $N_S$ are the total number of
particles in the Mott and superfluid phases, respectively. As seen above, in
equilibrium, (described by the configuration $f^z_0({\bf r})$), the
Mott-insulating particles can be thought of as providing an effective potential
that confines the superfluid particles. As appropriate to Josephson physics,
one can consider a situation in which the superfluid region slightly shrinks or
enlarges from its equilibrium configuration by transferring a small number of
particles $\delta N$ to or from the Mott region. In this situation, the bulk
energy takes the form
\begin{eqnarray}
\label{eq:bulk_deviation}
E_B & \approx & E_B^{Mott}(N_{M0})+E_B^{sf}(N_{S0}) \nonumber \\
& & + \frac{1}{2}\left(\left. \frac{\partial^2E_B^{Mott}}{\partial
N_M^2}\right|_0+ \left. \frac{\partial^2E_B^{sf}}{\partial
N_S^2}\right|_0\right)(\delta N)^2,
\label{eq:E_B}
\end{eqnarray}
where the subscript `$0$' denotes equilibrium. The energy scale for transfer of
particles to the superfluid region, $E_C$ (often called the ``capacitive energy"
in reference to Josephson physics in mesoscopic superconductors) is defined by
$E_{B} = E_C (\delta N)^2 /2$ and can be explicitly calculated from
Eq.(\ref{eq:E_B}) for a given external potential $V({\bf r})$. We observe that
Eq.(\ref{eq:E_B}) implies that the bulk-energy depends quadratically, rather than linearly, on the number of particles transfered. This differs from the result
for two externally-trapped superfluids, where the
linear contribution to the bulk energy only vanishes when the energy change of
the two coupled superfluid regions is combined\cite{Zapata98}. In the
superfluid-Mott coexisting phases, the transfer of particles is a local one
between one superfluid region and the surrounding Mott phase, whose boundary is
determined by the external potential and the ratio $J/U$ and is in this sense
self-organized. The coexisting system thus obeys an equilibrium condition
between the Mott and superfluid regions: $\left. \partial E_B^{Mott}/\partial
N_M \right|_0= \left. \partial E_B^{sf}/\partial N_S \right|_0$, rendering the
lowest order dependence on $\delta N$ quadratic as opposed to linear.

Turning to the dynamics governing the Mott-superfluid system, in the
pseudo-spin approximation the spins obey Heisenberg equations of motion.
Furthermore, in the mean-field approximation, the spins obey Bloch equations,
$\partial_t{\bf f}_i = {\bf f}_i \times {\bf B}_i$. To properly capture the Josephson coupling between superfluid regions, we go beyond the Thomas-Fermi approximation and allow
spatial variations in the spin operators: $\sum_j {\bf f}_j\approx z {\bf
f}+a^2\nabla^2{\bf f}$ in the continuum approximation. The resulting equations
of motion for the local order parameter can be used to derive the collective
modes within each superfluid region\cite{Barankov07}. The equation of
motion for the $z$-component of the spin system can be written in the form of a
continuity equation, $\partial_t \langle n \rangle + {\nabla} \cdot \vec{{\cal
J}} = 0$, where
\begin{equation}
\label{eq:current}
\vec{{\cal J}} = i J a^2(\psi\vec{\nabla}\psi^*-\psi^*\vec{\nabla}\psi)
\end{equation}
can be identified as the supercurrent density of bosons. To calculate the Josephson
coupling between spatially-separated superfluid regions, we note that close to the Mott-superfluid interface, the Thomas-Fermi approximation breaks down and the order parameter respects the equation (for $f^z <0$, i.e. the boundary with the $n_0$-Mott region)
\begin{eqnarray}
\label{eq:psop}
i\partial_t\psi & \approx & - J(n_0+1)a^2\nabla^2\psi \nonumber
\\ && +[Un_0-\tilde{\mu}-zJ(n_0+1)]\psi+2Jz|\psi|^2\psi.
\end{eqnarray}
(A similar equation is respected close to the $n_0+1$-Mott boundary where $f^z
> 0$.) This equation can be used to find the spatial decay of the order parameter
beyond the Thomas-Fermi boundary of the Mott region. We remark that this
Gross-Pitaevskii-type dynamics of the order parameter\cite{Pethick_Smith}
breaks down well within the condensed phase. In particular, as seen above in
Eqns.(\ref{eq:OP}) and (\ref{eq:density}), the Mott-superfluid system does not
have a density of bosons directly proportional to the square of the order
parameter.

While the pseudo-spin description suffices at the Mott-superfluid boundary, and
is in fact ideally suited to connect the magnitude of the order parameter at
the boundary to its value in the bulk of the superfluid, it does not capture
the physics deep in the Mott region between superfluid regions. In the
$n_0$-Mott region, for instance, it is clear that we need to consider
occupation numbers $n_0+1$ and $n_0-1$ in addition to $n_0$.  The relevant
equations of motion for this case are easily calculated by employing a mean
field perturbation theory\cite{Sachdev_book}, and one finds (given here to
lowest non-vanishing order in $\psi$):
\begin{eqnarray}
\label{eq:ftop}
i\kappa_{\tau} \partial_{t} \psi& \approx & - \kappa_r {\bf
\nabla}^2\psi+\alpha \psi, \nonumber \\
\alpha & = & \frac{1}{a^3}\left[\frac{1}{zJ}-\frac{n_0+1}{Un_0-\tilde{\mu}}
-\frac{n_0}{\tilde{\mu}-U(n_0-1)}\right],
\end{eqnarray}
where $\kappa_{\tau}=a^{-3} \frac{\partial\alpha}{\partial\mu}$ and
$\kappa_r=\frac{a^{-1}}{z^2J}$. At the mean-field level, the Mott-superfluid
boundary is captured by the relationship $\alpha=0$, which can be used to
generate the Mott lobes of the Bose-Hubbard phase diagram shown in
Fig.\ref{fig:domains}. Furthermore, the equations of motion obtained by this
approach, as required, coincide with Eq.(\ref{eq:psop})
close to the superfluid boundary (where terms of order $|\psi|^3$ can be
ignored).

We are now equipped to derive the Josephson coupling between two superfluid
regions $A$ and $B$, described by corresponding order parameters
$\psi_Ae^{i\phi_A}$ and $\psi_Be^{i\phi_B}$, where $\psi_{A/B}$ are real.
Assuming a total order parameter of the form
$\psi_Ae^{i\phi_A}+\psi_Be^{i\phi_B}$,  the continuity equations of the two
superfluid regions combine to give a continuity equation between the two
regions: $\partial_t (\langle n \rangle_A - \langle n\rangle_B) + \nabla \cdot
\vec{\cal J} = 0$, where $\vec{\cal J}$ given by Eq.(\ref{eq:current}), is
found to have the Josephson form:
\begin{eqnarray}
\label{eq:josephson}
\vec{{\cal J}} = 2J
a^2\left(\psi_B\vec{\nabla}\psi_A-\psi_A\vec{\nabla}\psi_B\right)\sin(\phi_{AB}),
\end{eqnarray}
where $\phi_{AB}=\phi_A-\phi_B$ is the relative phase between the superfluids.
The Josephson energy is defined by $\partial_t (\delta N_{A\rightarrow B}) =
-E_J \sin(\phi_A-\phi_B)$, where when particles are transferred from the A
region to the B region, $\delta N_A = -\delta N_B = \delta N_{A\rightarrow B}$.
$E_J$ can be explicitly calculated from
Eq.(\ref{eq:josephson}) and the above continuity equation by integrating over
an appropriate surface enclosing one of the superfluid regions. One finds that
 $E_J$ is proportional to the overlap of the order parameters $\psi_A$ and
$\psi_B$ in the region separating the two superfluids. In the two situations
depicted in Fig. \ref{fig:domains}, this Josephson coupling a) behaves as a
weak link bridging the two superfluid domains along the line of closest
approach or b) has a radially symmetric form connecting two concentric
superfluid shells, and its evaluation can be reduced to a one-dimensional
problem along the appropriate direction. In fact, the equilibrium configuration
given by Eq. (\ref{eq:ftop}) has a direct correspondence with the
Ginzburg-Landau form for superconductors\cite{Tinkham_book} and with the
Gross-Pitaevksii form for a superfluid\cite{Pethick_Smith}  trapped in a
potential, given in this case by $\alpha({\bf r})$.
Hence, we can use standard techniques for calculating the Josephson coupling
for a one-dimensional system\cite{Dalfovo96,Zapata98} and by employing the WKB
approximation for the superfluid order parameters in the Mott region, we find
\begin{equation}
\label{eq:Josephson}
E_J \approx A_J \exp\left[-\int_C\sqrt{Q({\bf r'})}d{\bf r}'\right],
\end{equation}
where $Q({\bf r'})= z^2Ja\alpha({\bf r'})$. The contour $C$ can be evaluated
using the method of steepest descent and is the least-action path linking the
two superfluids through the Mott-insulating barrier. Its end points correspond
to the two turning points at the Mott-superfluid interface for $A$ and $B$ at
which the function $\alpha$ vanishes. The constant $A_J$ depends on the precise
forms of $\psi_A$ and $\psi_B$. As in the case of condensates in free
space,\cite{Dalfovo96}, $A_J$ can be obtained by using a linearized potential
approximation and matching the boundary condition imposed at the
Mott-superfluid interface by Eq.(\ref{eq:psop}).
From Eq.(\ref{eq:Josephson}), a lower bound can be placed on the exponential
dependence of the Josephson coupling by setting $\alpha$ to its maximum value
of $1/(zJa^3)$ along the whole path $C$ to obtain a value of
$\exp(-\sqrt{z}{\ell}_{AB}/a)$, where ${\ell}_{AB}$ is the path length.
Strikingly, to first order, the Josephson coupling is dominated in an
exponential manner only by the path length between superfluid regions which in
turn is determined by the potential landscape. We remark that for the
Bose-Hubbard system, Eq.(\ref{eq:Josephson}) represents an explicit derivation
of the transport co-efficient postulated in Ref.\cite{Shimshoni98} on
phenomenological grounds.

To demonstrate the above formalism and to obtain estimates of the bulk and
Josephson energies for an already-realized experimental
system\cite{MottSfShells}, we now consider $N=10^6$ ultra-cold $^{87}$Rb atoms
in a three-dimensional optical lattice of spacing $a=0.43$~$\mu$m
(corresponding to a laser wavelength $\lambda = 2a$), hopping parameter
$J=h\times120$~Hz, and on-site repulsion $U=h\times10^4$~Hz confined by a
harmonic trap $V(r) = br^2$ with $b=h\times24$~Hz/$\mu\mbox{m}^2$. This system
has an inner Mott core with 2 atoms per site surrounded by a superfluid shell
(SFA), a Mott shell with $n=1$ atom per site (1-Mott) and finally an outer
superfluid shell (SFB); the Josephson coupling between the SFA and SFB shells
is mediated through the 1-Mott shell. Eq.(\ref{eq:ftop}) can be solved
for the locations where $\alpha =0$ to yield the boundaries of all the shells
in the system. To calculate the capacitive energy, $E_C$, one considers a
transfer of a small number of particles from SFA to SFB, which leads to a
change in the location of the regions' boundaries. Then,
Eq.(\ref{eq:bulk_energy}) can be used to find the change in energy of the
system. Linearizing the external potential in each superfluid region allows one
to obtain the following expression for the bulk energy for shell systems where
the coupling is through the $n$-Mott region:
$E_C =  (ba^2)^{3/2}/(6\pi \sqrt{U})[((2n+1)^2\sqrt{\mu/U-n})^{-1}
+((2n-1)^2\sqrt{\mu/U-(n-1)})^{-1}]$. For the parameters detailed above, this
leads to a bulk energy of $E_C \approx h\times 5 \times 10^{-3}$~Hz. The
Josephson energy can be calculated using Eq.(\ref{eq:josephson}) after solving
for the order parameter solutions near their respective boundaries using
Eq.(\ref{eq:psop}). These solutions each display a characteristic decay length,
$d_{A}=\sqrt[3]{J(n+1)a^2/q_{A}}$ and $d_{B}=\sqrt[3]{Jna^2/q_{B}}$,
respectively\cite{Dalfovo96}, where $q_{A/B} = dV/dr|_{r_{A/B}}$ is the slope
of the external potential at the boundary of each superfluid shell ($r_{A/B}$).
In terms of these quantities, the constant in Eq.(\ref{eq:Josephson}) takes the
form $A_J = (\pi J A^2/z) \sqrt{n(n+1)}(r_Ar_Ba)/(d_Ad_B)^{3/2}$, where $A
\approx0.397$\cite{Dalfovo96}. After a numerical integration, for the
parameters detailed above we find $E_J \approx A_J e^{-28} \approx h\times 2
\times 10^{-8}$~Hz. Because the Josephson energy is exponentially dependent on
the distance between the coupled superfluid regions, it may be possible to
obtain a significantly larger Josephson coupling in the case of a random (or
pseudo-random) external potential where this distance could be more easily
tuned. From the energies found above, we can predict that the shell system will
have Josephson oscillations which are in the strongly quantum regime ($E_J \ll
E_C$\cite{Zapata98}) and that the Josephson plasmon frequency, $\omega_{JP} =
\sqrt{E_JE_C} \sim 10^{-4}$~Hz is quite small. This suggests that the system
will be very slow (on the order of hours) to transfer particles between the two
shells and that a phase difference initially present between the superfluids
will remain for the duration of most current experiments, as can be ascertained
via interference experiments\cite{Andrews97}. Similar estimates in disordered
condensed matter systems, where Josephson physics is expected to play a major
role, are in order. Furthermore, a more complete description of such
co-existent Mott insulating and superfluid phases in inhomogeneous systems will
need to incorporate several relevant factors such as going beyond mean field
treatments, enlarging the truncated Hilbert space, and studying dissipative
effects, for instance, due to quasiparticle excitations.

We would like to acknowledge A. Auerbach and S. Sachdev for illuminating
discussions. This work was supported by the NSF under grants
DMR-0644022-CAR (SV) and DMR-0605871 (CL).

\end{document}